\documentclass[%
 aip,
 jcp,
 amsmath,amssymb,
 reprint,%
]{revtex4-1}

\usepackage{graphicx}
\usepackage{dcolumn}
\usepackage{bm}

\usepackage[utf8]{inputenc}
\usepackage[T1]{fontenc}
\usepackage{mathptmx}

\begin{document}

\preprint{AIP/123-QED}

\title{Bandgap of two-dimensional materials: Thorough assessment of modern exchange-correlation functionals}

\author{Fabien Tran}
\email{tran@theochem.tuwien.ac.at}
\affiliation{Institute of Materials Chemistry, Vienna University of Technology, Getreidemarkt 9/165-TC, A-1060 Vienna, Austria}
\author{Jan Doumont}
\affiliation{Institute of Materials Chemistry, Vienna University of Technology, Getreidemarkt 9/165-TC, A-1060 Vienna, Austria}
\author{Leila Kalantari}
\affiliation{Institute of Materials Chemistry, Vienna University of Technology, Getreidemarkt 9/165-TC, A-1060 Vienna, Austria}
\author{Peter Blaha}
\affiliation{Institute of Materials Chemistry, Vienna University of Technology, Getreidemarkt 9/165-TC, A-1060 Vienna, Austria}
\author{Tom\'{a}\v{s} Rauch}
\affiliation{Institut f\"ur Festk\"orpertheorie und -optik, Friedrich-Schiller-Universit\"at Jena and European Theoretical Spectroscopy Facility, Max-Wien-Platz 1, 07743 Jena, Germany}
\author{Pedro Borlido}
\affiliation{Institut f\"ur Festk\"orpertheorie und -optik, Friedrich-Schiller-Universit\"at Jena and European Theoretical Spectroscopy Facility, Max-Wien-Platz 1, 07743 Jena, Germany}
\author{Silvana Botti}
\affiliation{Institut f\"ur Festk\"orpertheorie und -optik, Friedrich-Schiller-Universit\"at Jena and European Theoretical Spectroscopy Facility, Max-Wien-Platz 1, 07743 Jena, Germany}
\author{Miguel A. L. Marques}
\affiliation{Institut f\"{u}r Physik, Martin-Luther-Universit\"{a}t Halle-Wittenberg, D-06099 Halle, Germany}
\author{Abhilash Patra}
\affiliation{School of Physical Sciences, National Institute of Science Education and Research, HBNI, 
Bhubaneswar 752050, India}
\author{Subrata Jana}
\affiliation{School of Physical Sciences, National Institute of Science Education and Research, HBNI, 
Bhubaneswar 752050, India}
\author{Prasanjit Samal}
\affiliation{School of Physical Sciences, National Institute of Science Education and Research, HBNI, 
Bhubaneswar 752050, India}

\begin{abstract}

The density functional theory (DFT) approximations that are the most accurate for the calculation of band gap of bulk materials are hybrid functionals like HSE06, the MBJ potential, and the GLLB-SC potential. More recently, generalized gradient approximations (GGA), like HLE16, or meta-GGAs, like (m)TASK, have proven to be also quite accurate for the band gap. Here, the focus is on 2D materials and the goal is to provide a broad overview of the performance of DFT functionals by considering a large test set of 298 2D systems. The present work is an extension of our recent studies [Rauch \textit{et al}., Phys. Rev. B \textbf{101}, 245163 (2020) and Patra \textit{et al}., J. Phys. Chem. C \textbf{125}, 11206 (2021)]. Due to the lack of experimental results for the band gap of 2D systems, $G_{0}W_{0}$ results were taken as reference. It is shown that the GLLB-SC potential and mTASK functional provide the band gaps that are the closest to $G_{0}W_{0}$. Following closely, the local MBJ potential has a pretty good accuracy that is similar to the accuracy of the more expensive hybrid functional HSE06.

\end{abstract}

\maketitle

\section{\label{sec:introduction}Introduction}

The calculation of the band gap with density functional theory \cite{HohenbergPR64,KohnPR65} (DFT) is computationally efficient, but can also be quite accurate, provided that a suitable approximation for the exchange-correlation (xc) effects is chosen. Particularly interesting are the semilocal xc functionals, which are the fastest to evaluate. The most used semilocal xc functionals for the band gap of bulk solids are the modified Becke-Johnson (MBJ) potential \cite{TranPRL09} and the GLLB-SC potential.\cite{GritsenkoPRA95,KuismaPRB10} The absolute errors with respect to experiment that are obtained with the hybrid functionals (e.g., HSE06 \cite{HeydJCP03,KrukauJCP06}), MBJ, and GLLB-SC are on average in the range 0.5$-$0.8~eV (15$-$40\% for the absolute relative error) depending on the test set.\cite{TranJPCA17,TranPRM18,TranJAP19,BorlidoJCTC19,BorlidoNPJCM20} Such errors are much lower than the average errors obtained with the standard PBE functional \cite{PerdewPRL96} that are in the range 1$-$2~eV (around 50\% for the relative error). Proposed recently, the HLE16 \cite{VermaJPCL17} generalized gradient approximation (GGA), and the HLE17,\cite{VermaJPCC17} MGGAC, \cite{PatraPRB19b} TASK,\cite{AschebrockPRR19} and modified TASK \cite{NeupanePRM21} (mTASK) meta-GGAs (MGGA) are examples of other fast semilocal DFT functionals that can be quite accurate for the band gap of solids as well (see also Refs.~\onlinecite{ArmientoPRL13,FinzelIJQC17,PatraPRB19,TancogneDejeanPRB20,JanaNJP21}).

Thus, there are fast semilocal xc functionals that are useful alternatives to the much more expensive hybrid functionals and quasiparticle $GW$ methods,\cite{HedinPR65} albeit the latter are more accurate if performed self-consistently and with vertex corrections.\cite{ShishkinPRL07,ChenPRB15} Actually, it should be mentioned that dielectric-dependent hybrid functionals are also more accurate than the hybrids using a fixed amount of exact exchange.\cite{KollerJPCM13,SkonePRB14,ChenPRM18,ZhangWCMS20}

The majority of benchmarks of xc functionals for the band gap are done using test sets composed of bulk solids. Furthermore, only bulk solids are typically used when free parameters in a xc functional are optimized.\cite{TranPRL09,FinzelIJQC17} Comparatively, there are much less similar studies on atomically-thin films, often referred to as two-dimensional (2D) systems. Such works on 2D systems related to the assessment of xc functionals for band gaps can be found in Refs.~\onlinecite{MiroCSR14,RasmussenJPCC15,BudaSR17,SchmidtPRB17,Haastrup2DM18,OlsenNPJCM19,PatraPCCP19,PatraPRB20,RauchPRB20,ZhuCPB20,DiSabatinoFD20,PatraJPCC21,NeupanePRM21}. We note that an important difference between bulk solids and 2D systems concerns the magnitude of the excitonic effect (present in optical measurements), which is an effect beyond the Kohn-Sham (KS) and $GW$ methods. It is usually small for most bulk solids (except wide-gap insulators), of the order of tens of meV, but can reach several eV for systems with reduced dimensionality (see, e.g., Ref.~\onlinecite{WirtzPRL06}). Therefore, a direct comparison between the theoretical quasiparticle band gap and the gap obtained from optical experiments should in principle not be done.

Alternatively, fundamental band gaps calculated from the $GW$ quasiparticle method can be used as reference, since $GW$ is viewed as the state-of-the-art for band structure calculations. In Refs.~\onlinecite{Haastrup2DM18,RauchPRB20,PatraPRB20,PatraJPCC21} for instance, $GW$ band gaps were used for the testing of various xc DFT functionals on 2D materials (see Ref.~\onlinecite{LeePRB16} for bulk solids). The goal of the present work is to provide a thorough assessment of DFT approximations for the band gap of 2D materials. It is a follow-up study of our recent works \cite{RauchPRB20,PatraJPCC21} devoted to a more systematic comparison of xc functionals. In particular some of the most recent MGGA functionals will be considered. Studies reporting benchmark tests of theoretical methods are especially important for 2D materials nowadays. \cite{RasmussenJPCC15,Haastrup2DM18,RauchPRB20,RasmussenNPJCM21,VandenbergheJCE21} Indeed, these systems can show exceptional properties that are not found in bulk materials and can be very relevant for technological applications.\cite{BhimanapatiACSN15}

The paper is organized as follows. In Sec.~\ref{sec:method} a brief description of the tested functionals and test set is given. Section~\ref{sec:results} discusses the results and Sec.~\ref{sec:summary} gives a summary of the conclusions.

\section{\label{sec:method}Methods and computational details}

We describe below the tested xc functionals. The Perdew-Burke-Ernzerhof (PBE) functional \cite{PerdewPRL96} is the standard GGA functional for solid-state total-energy calculations, however it is not accurate for band gaps. The GGA EV93PW91, which consists of the exchange from Engel and Vosko (EV93) \cite{EngelPRB93} and Perdew-Wang (PW91) correlation \cite{PerdewPRB92b} slightly improves over PBE for the band gap.\cite{TranJPCM07,TranJPCA17,BorlidoNPJCM20} Also considered is the high local exchange (HLE16) functional,\cite{VermaJPCL17} which is one of the most accurate GGAs for the band gap of bulk solids.\cite{TranJPCA17,BorlidoJCTC19} We mention that the Armiento-K\"{u}mmel (AK13) functional \cite{ArmientoPRL13} is another GGA that is good for the band gap of bulk solids, \cite{TranJPCA17,BorlidoNPJCM20} however it leads to numerical problems in 2D materials due to the presence of vacuum. For instance, no self-consistent-field (SCF) convergence could be achieved for many of the systems studied here. Therefore, AK13 will not be considered in the present work.

The MGGA functionals, which represent the next level of approximation,\cite{PerdewAIP01} can be more accurate. The ones that are tested here are r$^{2}$SCAN,\cite{FurnessJPCL20} which is a numerically more stable version of the well-known strongly constrained and appropriately normed (SCAN),\cite{SunPRL15} HLE17,\cite{VermaJPCC17} TASK \cite{AschebrockPRR19} and its modified version mTASK, \cite{NeupanePRM21} and MGGAC.\cite{PatraPRB19b} SCAN and r$^{2}$SCAN are general purpose functionals that were constructed by satisfying as many constraints as possible. Results for the band gap of bulk solids show that SCAN gives average errors similar to EV93PW91, i.e., the improvement with respect to PBE is moderate.\cite{BorlidoNPJCM20} However, for strongly correlated systems SCAN is better than EV93PW91.\cite{TranJPCA17,ZhangPRB20} Like HLE16, HLE17 was modelled specifically for band gaps and molecular excitation energies, and therefore yields quite accurate band gaps with errors very similar to HLE16.\cite{BorlidoNPJCM20} TASK was constructed non-empirically and also leads to accurate band gaps of bulk solids as shown in Refs.~\onlinecite{AschebrockPRR19,BorlidoNPJCM20}. However, TASK does not seem to be a general purpose functional like SCAN/r$^{2}$SCAN, since the lattice constants are very inaccurate.\cite{Doumont21} The same can probably be said about mTASK, which differs from TASK by the value of two parameters that were modified to make the enhancement factor more nonlocal and thus to increase the band gap.\cite{NeupanePRM21} MGGAC contains parameters, some of them were determined using mathematical constraints (e.g., uniform electron gas limit or the tight Lieb-Oxford bound), while others were fitted to the exchange energy of noble gas atoms or lattice constants of bulk solids. Within the general purpose functionals, MGGAC provides results close to HSE06.

All the xc approximations listed so far are based on an energy functional. This is not the case for GLLB-SC \cite{KuismaPRB10} and local MBJ (LMBJ) \cite{RauchJCTC20} that were modelled at the level of the xc potential and are not derivative of an energy functional. GLLB-SC is parameter-free and has been shown to be very accurate for the band gap of bulk solids, \cite{KuismaPRB10,CastelliEES12a,CastelliAEM14,PandeyJPCC17,TranPRM18,TranJAP19} although it does not perform as well as the MBJ potential as shown in Refs.~\onlinecite{TranPRM18,TranJAP19}. Compared to all other DFT approximations considered here, GLLB-SC differs in the way the band gap is calculated. While for the other functionals the band gap is calculated just as the difference between the conduction band minimum (CBM) and valence band maximum (VBM), for GLLB-SC it is calculated by adding a derivative discontinuity to the CBM$-$VBM difference.\cite{KuismaPRB10} GLLB-SC band gaps have been calculated for 2D materials in the computational 2D materials database (C2DB) database, \cite{RasmussenJPCC15,Haastrup2DM18,Gjerding2DM21} and it was shown that the agreement with the $G_{0}W_{0}$ (one-shot $GW$ \cite{HybertsenPRB86}) band gaps is excellent.\cite{Haastrup2DM18} A similar conclusion was drawn in our recent work where selected 2D systems were considered. \cite{PatraJPCC21}

The LMBJ potential is an adaptation of the MBJ potential for systems with vacuum (molecules, thin films, surfaces) and interfaces.\cite{RauchJCTC20} MBJ depends on the average of $\left\vert\nabla\rho\right\vert/\rho$ over the unit cell ($\rho$ being the electron density), a quantity that is meaningful in periodic bulk solids, but not in case of supercells including vacuum and for interfaces. Instead, LMBJ uses a local average of $\left\vert\nabla\rho\right\vert/\rho$ [that is slightly modified, see Eq.~(\ref{eq:g})] so that it can be used for any kind of system. The LMBJ potential is given by
\begin{equation}
 v_{x}^{\text{LMBJ}}(\bm{r})=c(\bm{r})v_{x}^{\text{BR}}(\bm{r})+(3c(\bm{r})-2)\frac{1}{\pi}\sqrt{\frac{5}{6}}\sqrt{\frac{\tau(\bm{r})}{\rho(\bm{r})}},
 \label{eq:vxLMBJ}
\end{equation}
where $v_{x}^{\text{BR}}$ is the Becke-Roussel potential,\cite{BeckePRA89} $\tau$ is the kinetic-energy density, and $c$ is given by
\begin{equation}
c(\bm{r})=\alpha+\beta \bar{g}(\bm{r}).
\label{eq:c}
\end{equation}
In Eq.~(\ref{eq:c})
\begin{equation}
 \bar{g}(\bm{r})=\frac{1}{(2\pi\sigma^2)^{3/2}}\int g(\bm{r}')e^{-\frac{|\bm{r}-\bm{r}'|^2}{2\sigma^2}}d^3r',
 \label{eq:gbar}
\end{equation}
where
\begin{equation}
g(\bm{r})=\frac{1-\alpha}{\beta}\left [1-\text{erf}\left(\frac{\rho(\bm{r})}{\rho_{\text{th}}} \right) \right ]+\frac{|\nabla 
\rho(\bm{r})|}{\rho(\bm{r})}\text{erf}\left(\frac{\rho(\bm{r})}{\rho_{\text{th}}} \right).
\label{eq:g}
\end{equation}
The values of the parameters in Eqs.~(\ref{eq:c})-(\ref{eq:g}) chosen by Rauch \textit{et al}. (see erratum of Ref.~\onlinecite{RauchPRB20}) are $\alpha=0.488$, $\beta=0.5$~bohr, $\sigma=3.78$~bohr, and $\rho_{\text{th}}=6.96\times10^{-4}$~e/bohr$^{3}$. $\alpha$ and $\beta$ are originally from Ref.~\onlinecite{KollerPRB12}, while $\sigma$ is the smearing parameter that determines the size of the region over which the average of $g$ is done. Finally, $\rho_{\text{th}}$ is the threshold density, which corresponds to a Wigner-Seitz radius $r_{s}^{\text{th}}=\left((4/3)\pi\rho_{\text{th}}\right)^{-1/3}=7$~bohr.

A technical aspect of the LMBJ potential and its parents BJ \cite{BeckeJCP06} and MBJ \cite{TranPRL09} as well as AK13 \cite{ArmientoPRL13} should be mentioned. As discussed in Refs.~\onlinecite{BeckeJCP06,AschebrockPRB17a,AschebrockPRB17b}, these potentials do not tend to zero in the region far from the nuclei, but rather to a system-dependent constant. Since it is customary to set any xc potential to zero or to the LDA value where $\rho$ is very low (below some chosen density threshold) to avoid numerical instabilities, one has to be careful not to do it in a region of space where the CBM extends. Otherwise, the CBM will be artificially shifted to a wrong energy, which would also lead to a wrong band gap. Thus, one has to use a density threshold that is small enough to avoid this problem, but also not too small to avoid numerical instabilities. A value of $10^{-9}$ e/bohr$^{3}$ seems appropriate for LMBJ.

The LMBJ potential was tested in Ref.~\onlinecite{RauchPRB20} on the 2D materials of the C2DB database; in that work it was shown that LMBJ is almost as accurate as HSE06 in reproducing the $G_{0}W_{0}$ band gaps. In the present work, two sets of LMBJ band gaps will be shown and discussed. One set was obtained with the parameters from Rauch \textit{et al}. listed above, while the other one was obtained with $\beta=0.6$, which leads to results which agree better with $G_{0}W_{0}$.

The potential LB94 of van Leeuwen and Baerends,\cite{vanLeeuwenPRA94} which is also not a functional derivative, will be tested as well. Note that the correlation in LB94 is LDA.\cite{PerdewPRB92a}

While hybrid functionals are in general much more accurate than the standard GGA PBE
for the band gap of solids (see Refs.~\onlinecite{JanaJCP20,JanaPRB20,WangPNAS20} for
recent works), we will consider only
the well-known HSE06, since the focus of the present work is on the fast semilocal
functionals. However, this should have no impact on the conclusion of our work, since
previous benchmark studies\cite{BorlidoJCTC19,BorlidoNPJCM20,WangPNAS20} have shown that
HSE06 is already among the very best hybrid functionals (excluding those which are dielectric-dependent) for the band gap of solids.

The test set of 2D materials considered in the present work is the same as the one used by Rauch \textit{et al}.\cite{RauchPRB20} It consists of 298 nonmagnetic systems taken from the C2DB database. \cite{Haastrup2DM18} For all of them, $G_{0}W_{0}$ band gaps calculated with the GPAW code \cite{EnkovaaraJPCM10} are available and will be used as reference values. For 2D materials, experimental band gaps with excitonic effect subtracted are scarce in the literature. For a few systems, comparison of the $G_{0}W_{0}$ band gap with the one inferred from experiment shows that the agreement is pretty good.\cite{SchmidtPRB17,Haastrup2DM18} However, it may well be that some of the $G_{0}W_{0}$ values that we use as reference are not as accurate for various reasons, in particular because of the dependency on the input orbitals and eigenvalues (PBE was used as the reference ground-state functional.\cite{Haastrup2DM18}) Nevertheless, due to the large number of systems, we believe that possible inaccuracies in the $G_{0}W_{0}$ data should be small for the statistics and conclusions.

The SCF calculations with all xc functionals were done with the WIEN2k code,\cite{WIEN2k,BlahaJCP20} which is based on the augmented-plane-wave plus local orbitals method.\cite{Singh,KarsaiCPC17} Spin-orbit coupling was included for all systems. Parameters like the basis-set size and number of $\textbf{k}$-points in the Brillouin zone were chosen to be large enough so that the band gap is converged to within a few 0.01~eV. The self-consistent implementation of MGGA functionals in WIEN2k is very recent \cite{Doumont21} and uses the subroutines from the library of exchange-correlation functionals Libxc.\cite{LehtolaSX18,MarquesCPC12}

\section{\label{sec:results}Results and discussion}

\begin{table*}
\caption{\label{tab:band_gap}Summary statistics for the error in the DFT band gaps with respect to $G_{0}W_{0}$ reference values for the set of 298 2D materials. M(P)E, MA(P)E, and S(P)D denote the mean (percentage) error, mean absolute (percentage) error, and standard (percentage) deviation, respectively. $a$ and $b$ are the coefficients of the linear fit $y=ax+b$ (shown in Fig.~\ref{fig:band_gap}), $r$ is the Pearson correlation coefficient, and the last column is for the number of materials wrongly described as metallic. The type of approximation of the functionals is indicated in parenthesis (GLLB-SC depends on the eigenvalues $\epsilon_{i}$ and the exchange derivative discontinuity $\Delta_{x}$ is added to $\text{CBM}-\text{VBM}$). The units of the ME, MAE, SD, and $b$ are eV.}
\begin{ruledtabular}
\begin{tabular}{lcccccccccc}
Functional & ME & MAE & MPE & MAPE & SD & SPD & $a$ & $b$ & $r$ & False metals \\
\hline
HSE06\footnotemark[1] (hybrid) & -0.71 & 0.78 & -15 & 29 & 0.68 & 65 & 0.73 & 0.15 & 0.98 & 0 \\
PBE (GGA) & -1.49 & 1.50 & -48 & 51 & 0.96 & 32 & 0.57 & -0.16 & 0.98 & 0 \\
EV93PW91 (GGA) & -1.29 & 1.29 & -41 & 42 & 0.92 & 16 & 0.60 & -0.02 & 0.98 & 0 \\
HLE16 (GGA) & -1.05 & 1.05 & -35 & 39 & 0.69 & 41 & 0.73 & -0.21 & 0.98 & 3 \\
LB94 ($\sim$GGA) & -1.67 & 1.67 & -55 & 62 & 0.98 & 71 & 0.60 & -0.39 & 0.95 & 7 \\
r$^{2}$SCAN (MGGA) & -1.18 & 1.18 & -37 & 39 & 0.82 & 22 & 0.64 & -0.06 & 0.98 & 0 \\
HLE17 (MGGA) & -1.07 & 1.08 & -35 & 38 & 0.72 & 24 & 0.71 & -0.16 & 0.98 & 1 \\
MGGAC (MGGA) & -0.82 & 0.86 & -14 & 37 & 0.76 & 146 & 0.69 & 0.14 & 0.97 & 0 \\
TASK (MGGA) & -0.65 & 0.66 & -18 & 25 & 0.54 & 39 & 0.81 & -0.06 & 0.98 & 0 \\
mTASK (MGGA) & -0.48 & 0.52 & -13 & 21 & 0.50 & 35 & 0.86 & -0.05 & 0.98 & 0 \\
GLLB-SC ($\sim$LDA/GGA+$\epsilon_{i}$+$\Delta_{x}$) & -0.20 & 0.42 & -7 & 21 & 0.55 & 52 & 1.06 & -0.38 & 0.97 & 1 \\
LMBJ($\beta=0.5$) ($\sim$MGGA) & -0.73 & 0.78 & -18 & 35 & 0.56 & 84 & 0.82 & -0.17 & 0.98 & 0 \\
LMBJ($\beta=0.6$) ($\sim$MGGA) & -0.32 & 0.50 & 1 & 32 & 0.60 & 155 & 0.92 & -0.07 & 0.96 & 0 \\
\end{tabular}
\end{ruledtabular}
\footnotetext[1]{GPAW results from Ref.~\onlinecite{Haastrup2DM18}.}
\end{table*}

\begin{figure*}
\includegraphics{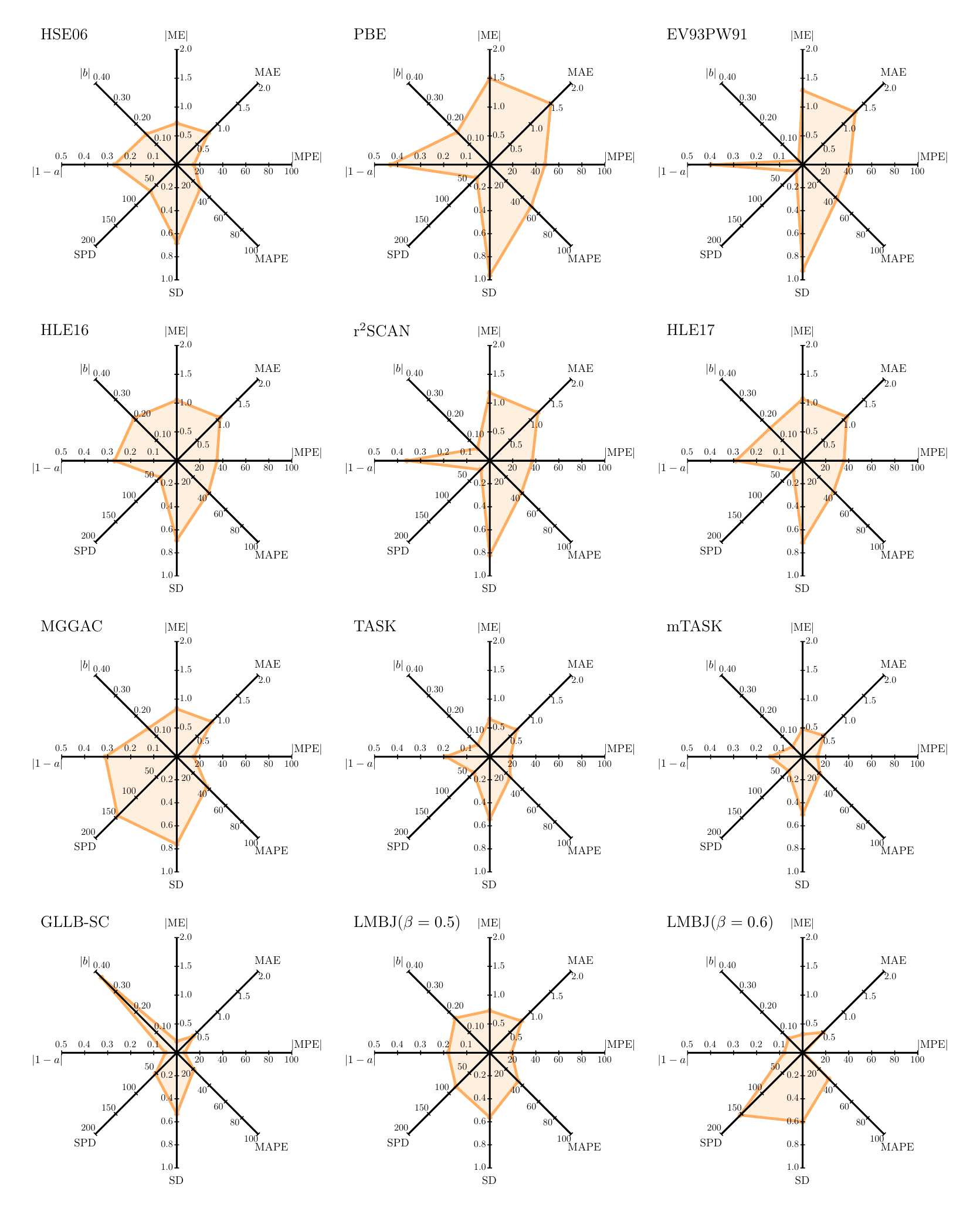}
\caption{\label{fig:radar}Radar plots showing the statistical quantities in Table~\ref{tab:band_gap} for all xc functionals except LB94.}
\end{figure*}

\begin{figure*}
\includegraphics[scale=0.5]{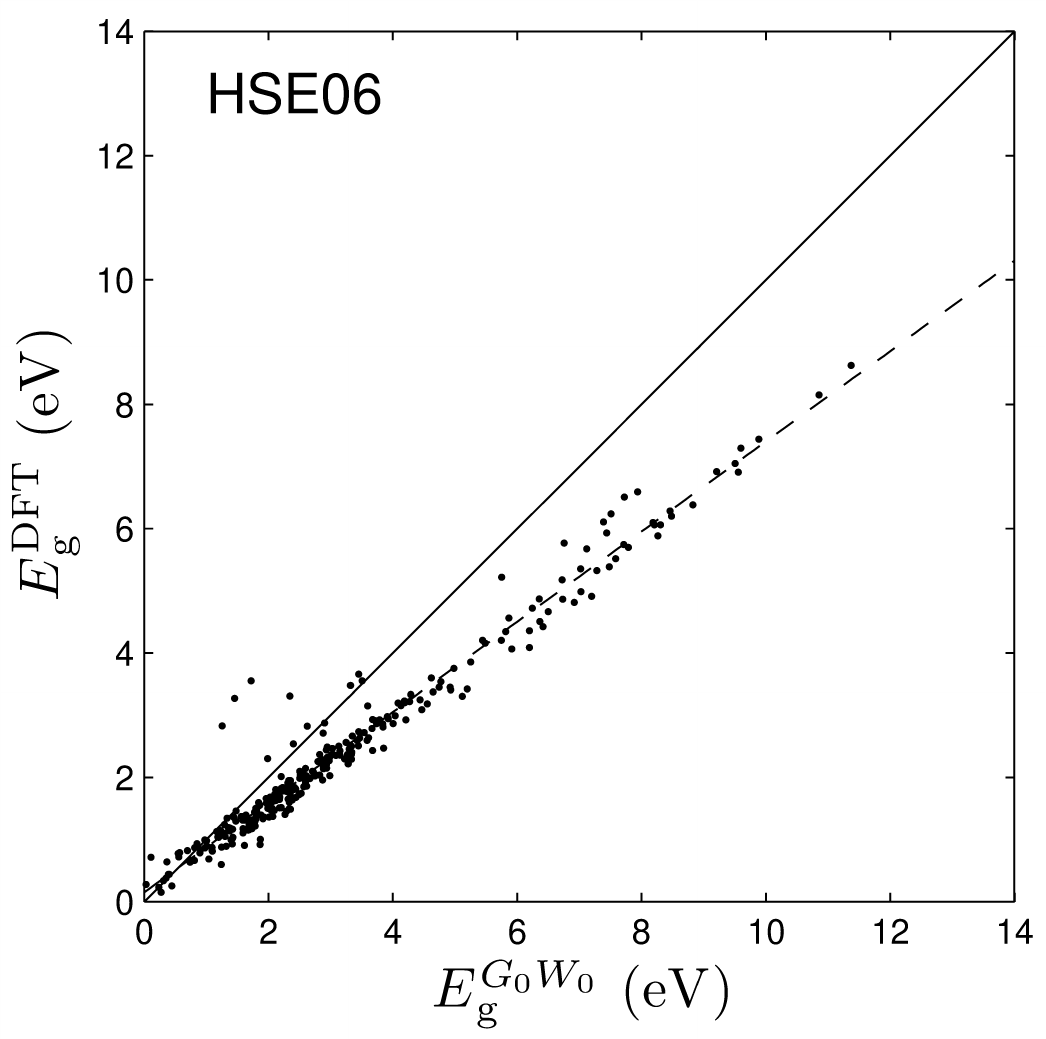}
\includegraphics[scale=0.5]{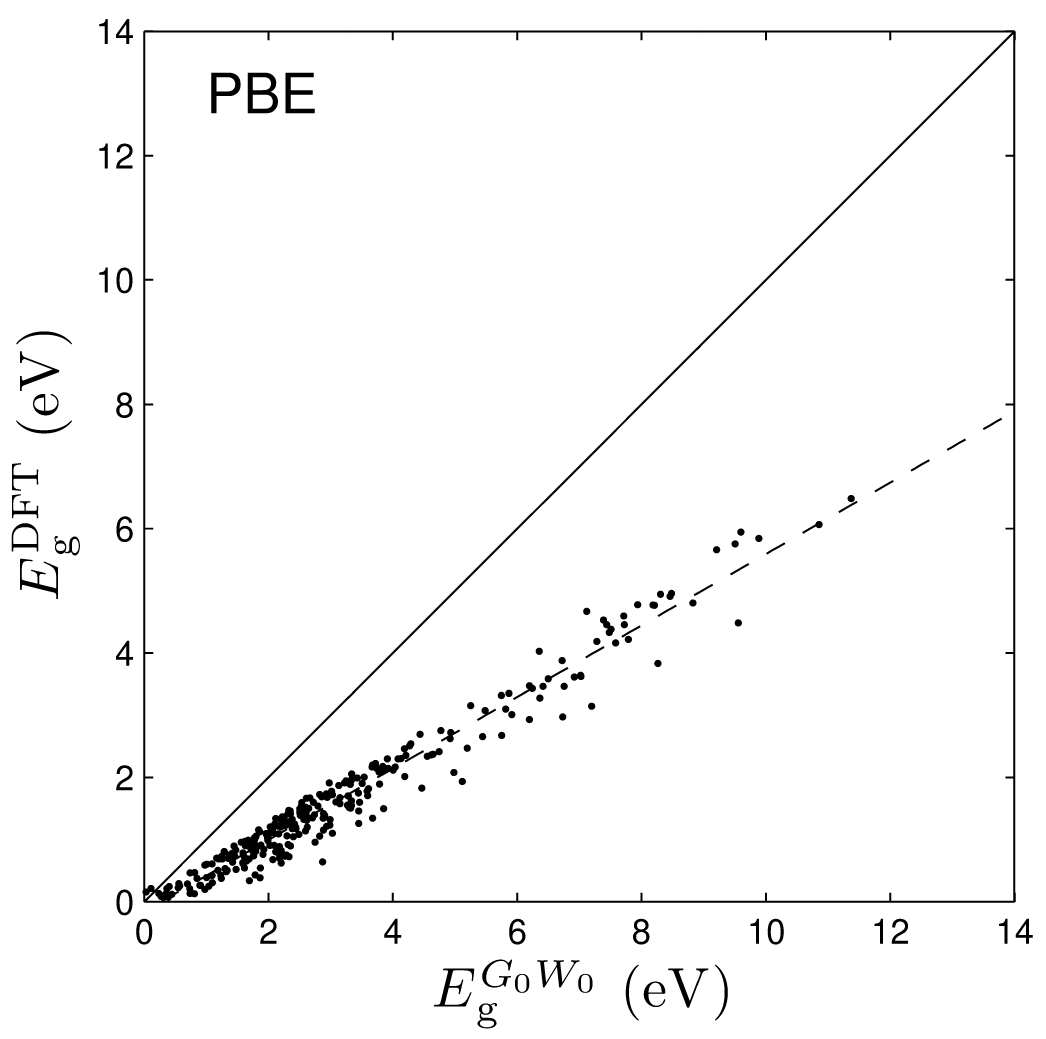}
\includegraphics[scale=0.5]{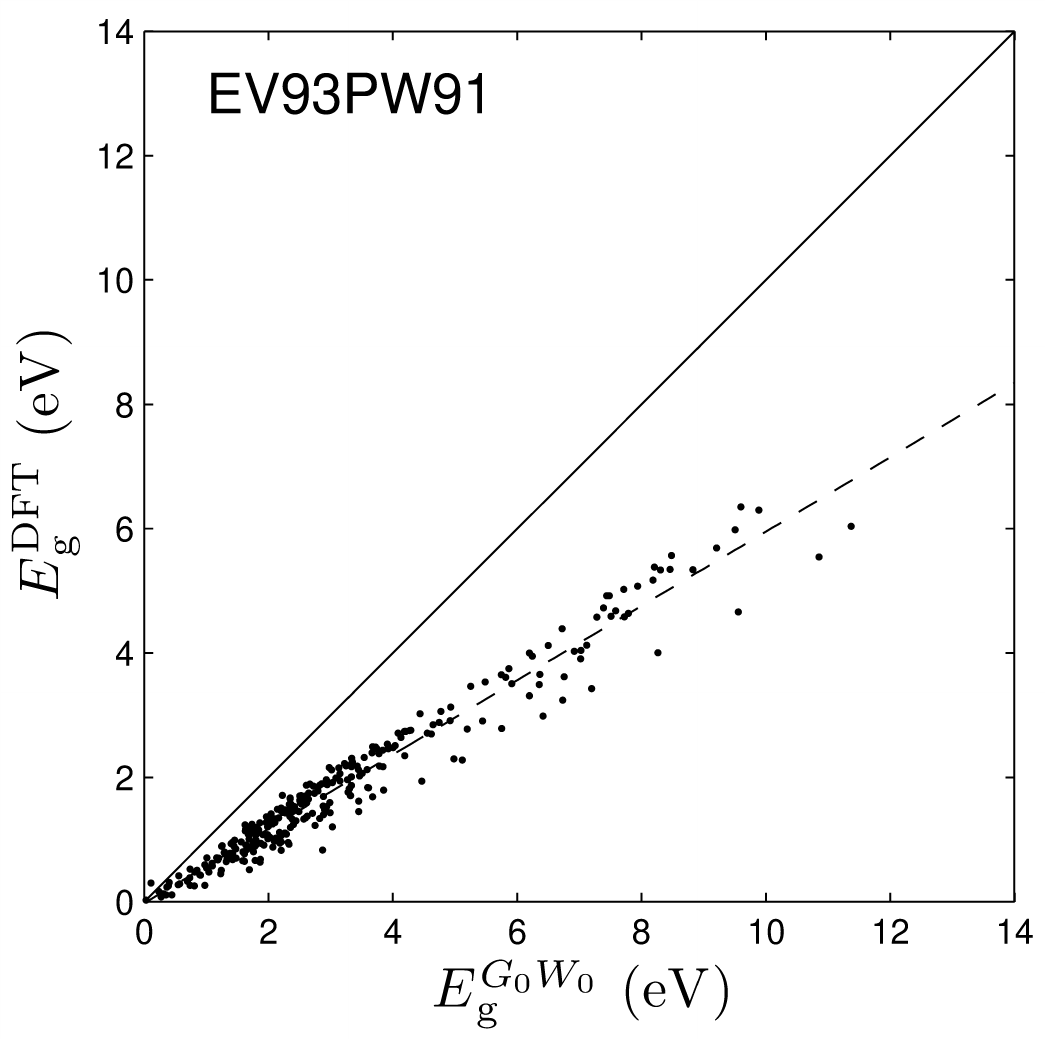}
\includegraphics[scale=0.5]{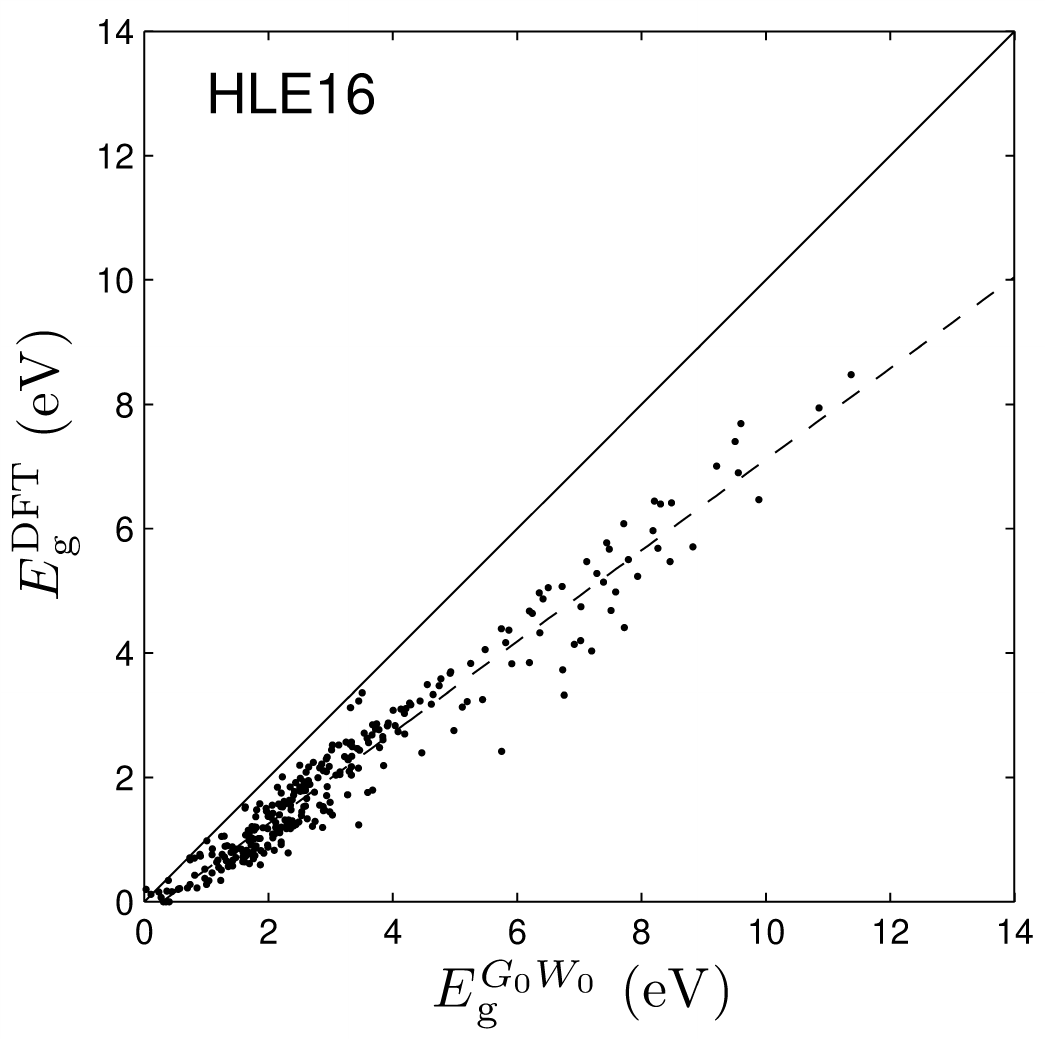}
\includegraphics[scale=0.5]{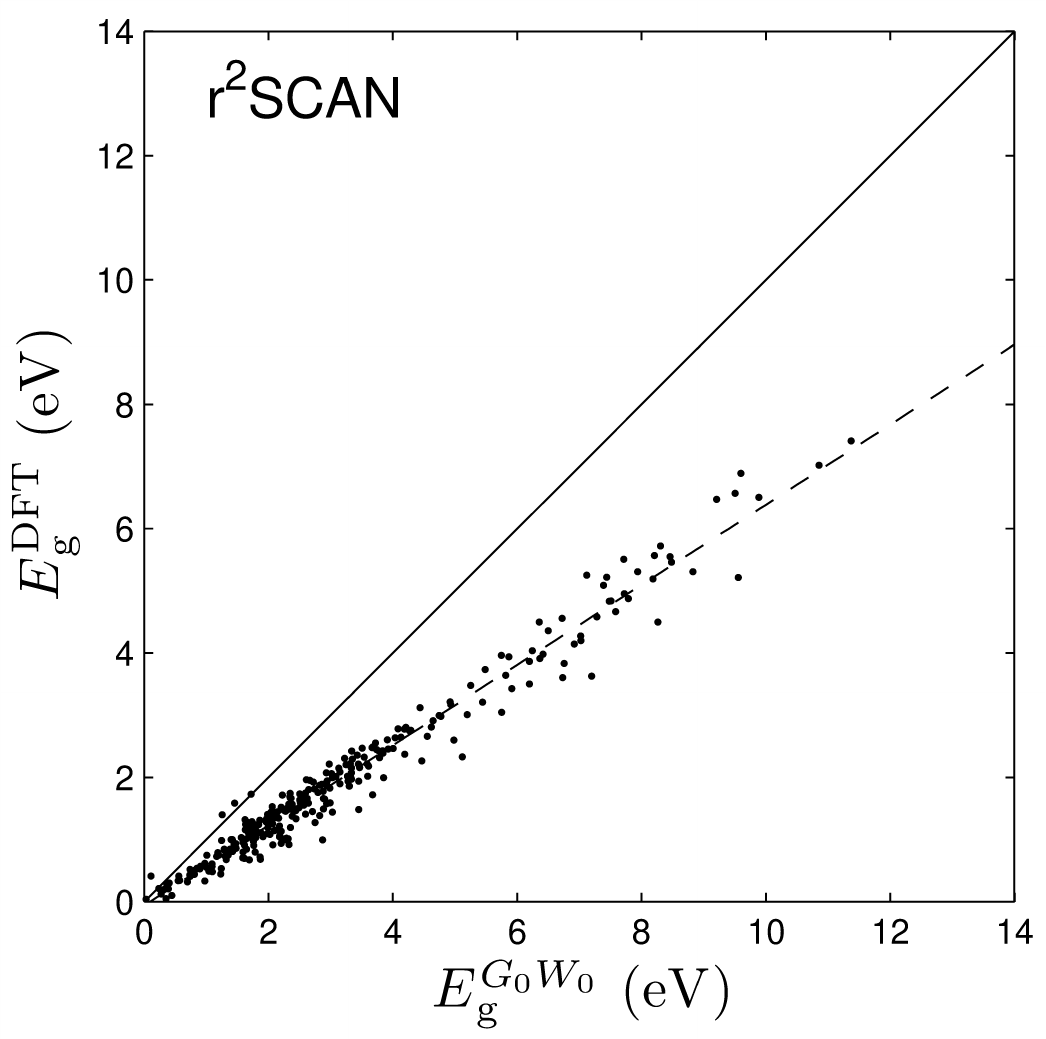}
\includegraphics[scale=0.5]{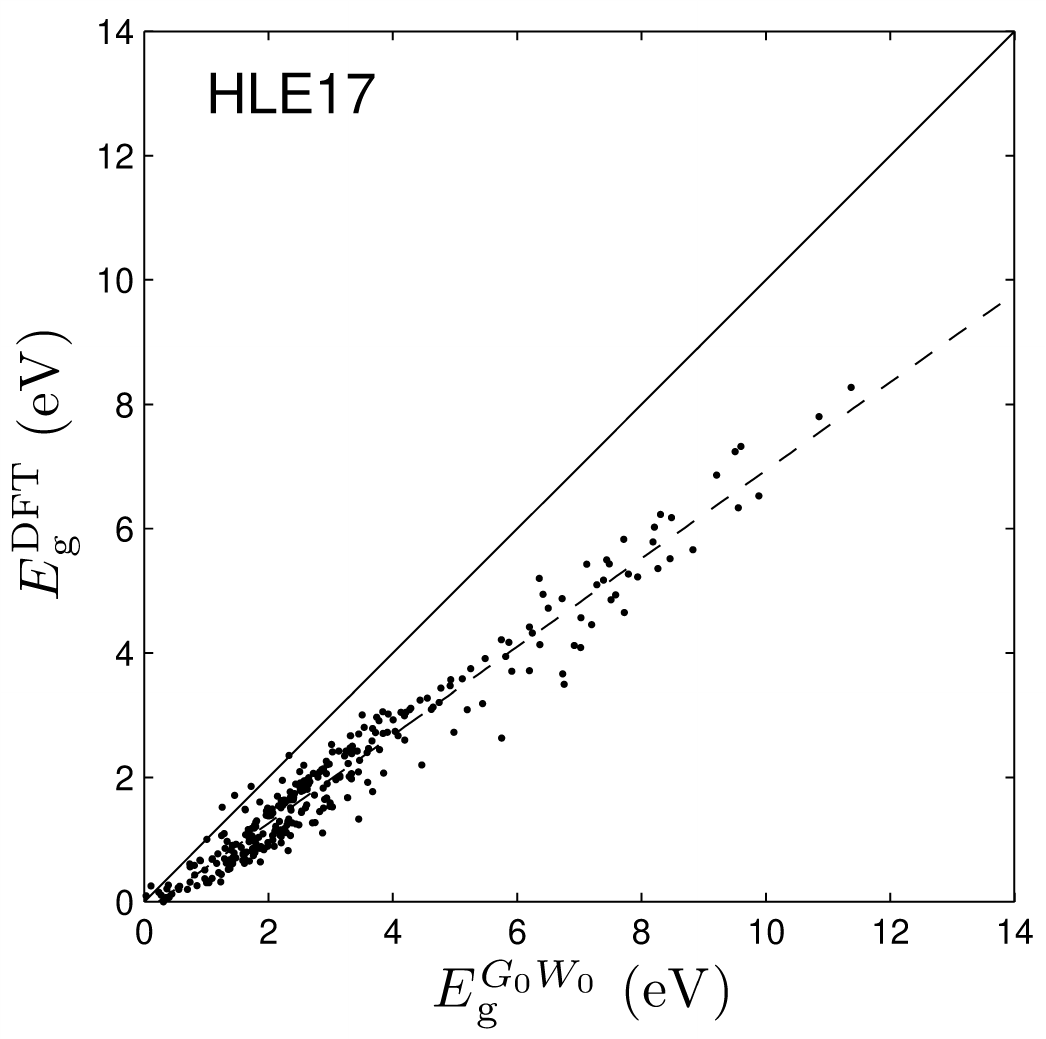}
\includegraphics[scale=0.5]{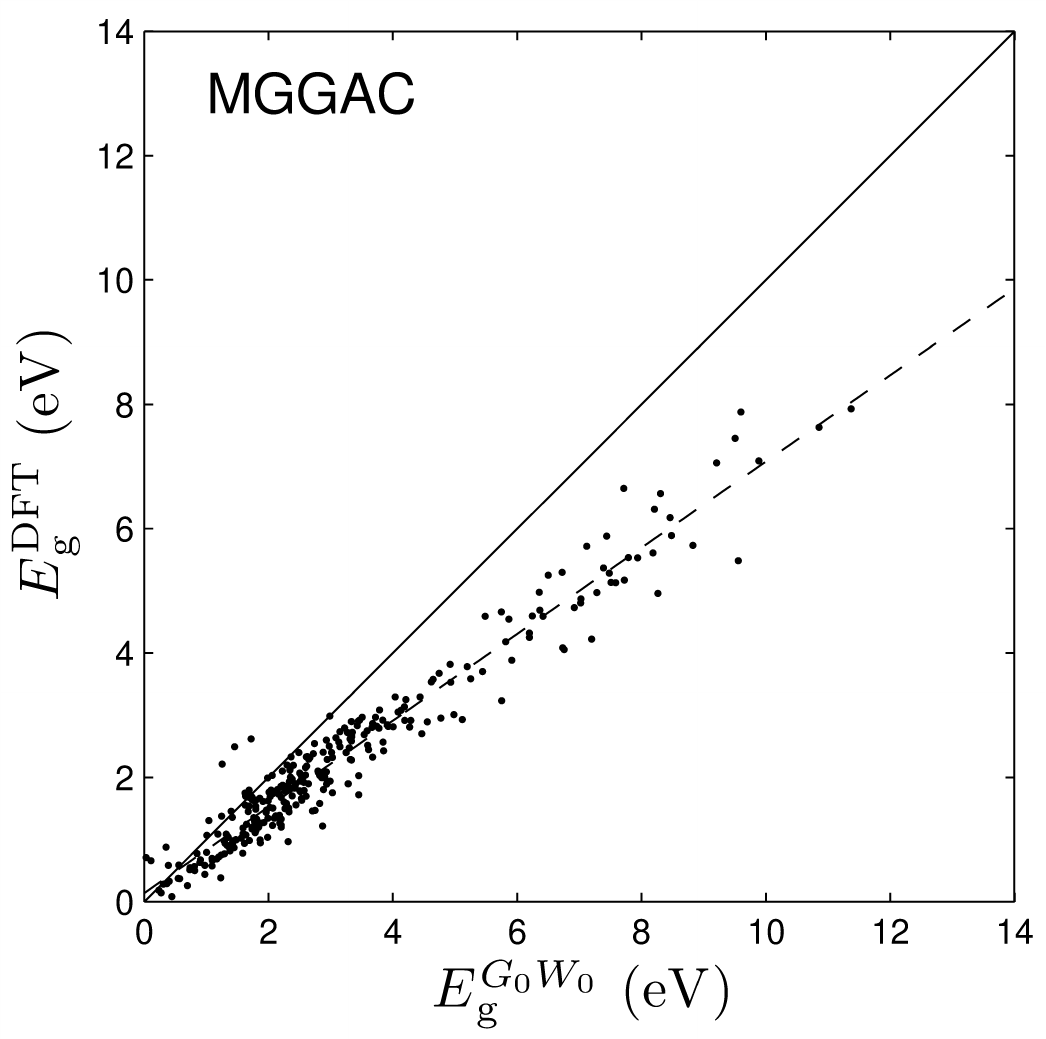}
\includegraphics[scale=0.5]{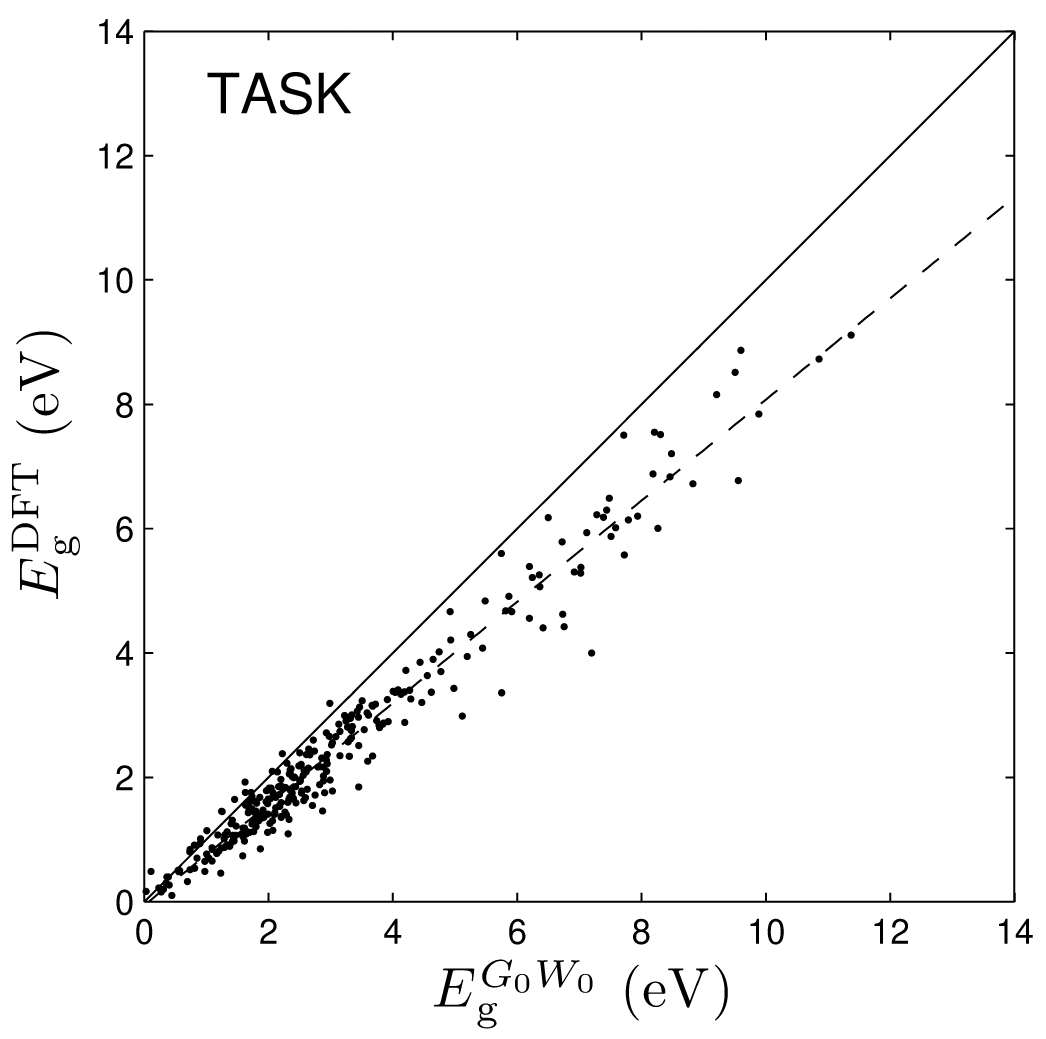}
\includegraphics[scale=0.5]{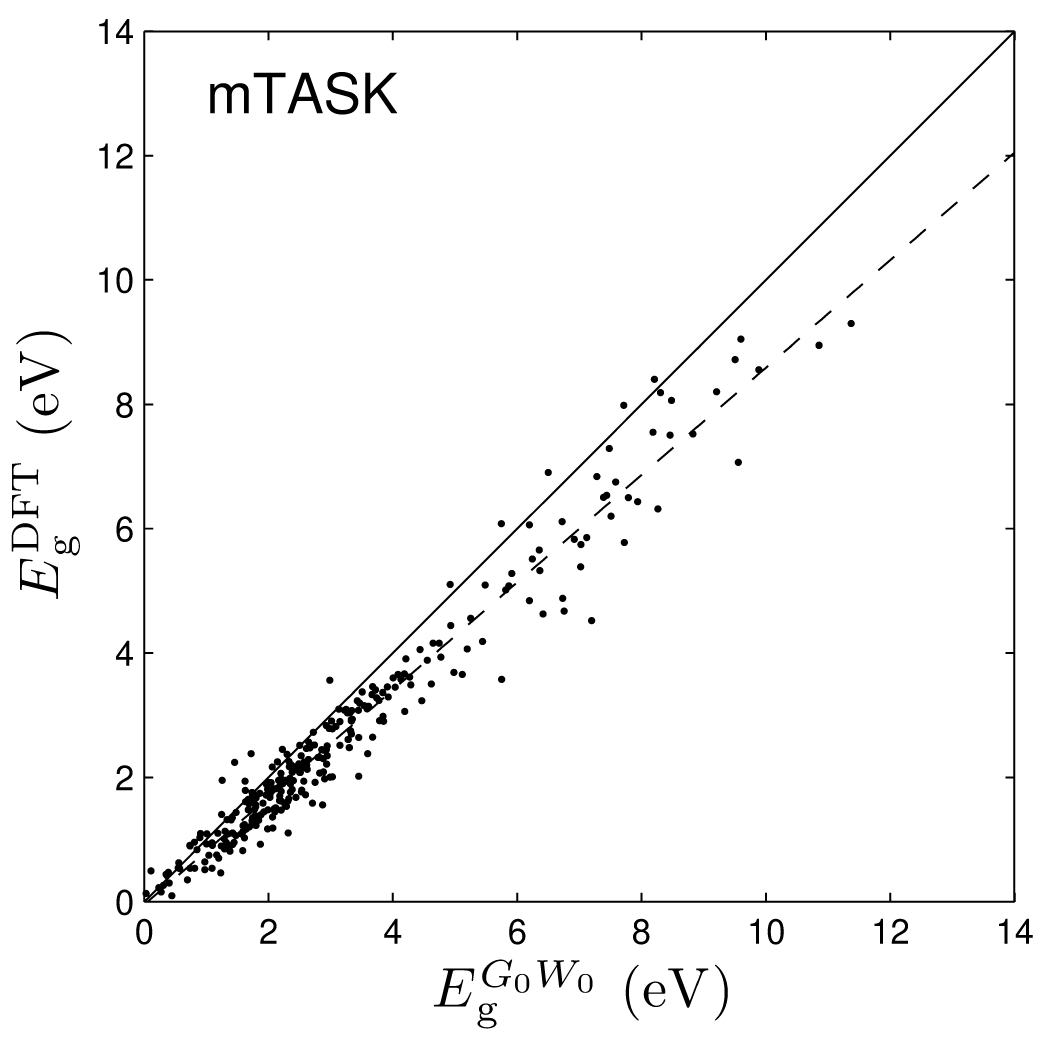}
\includegraphics[scale=0.5]{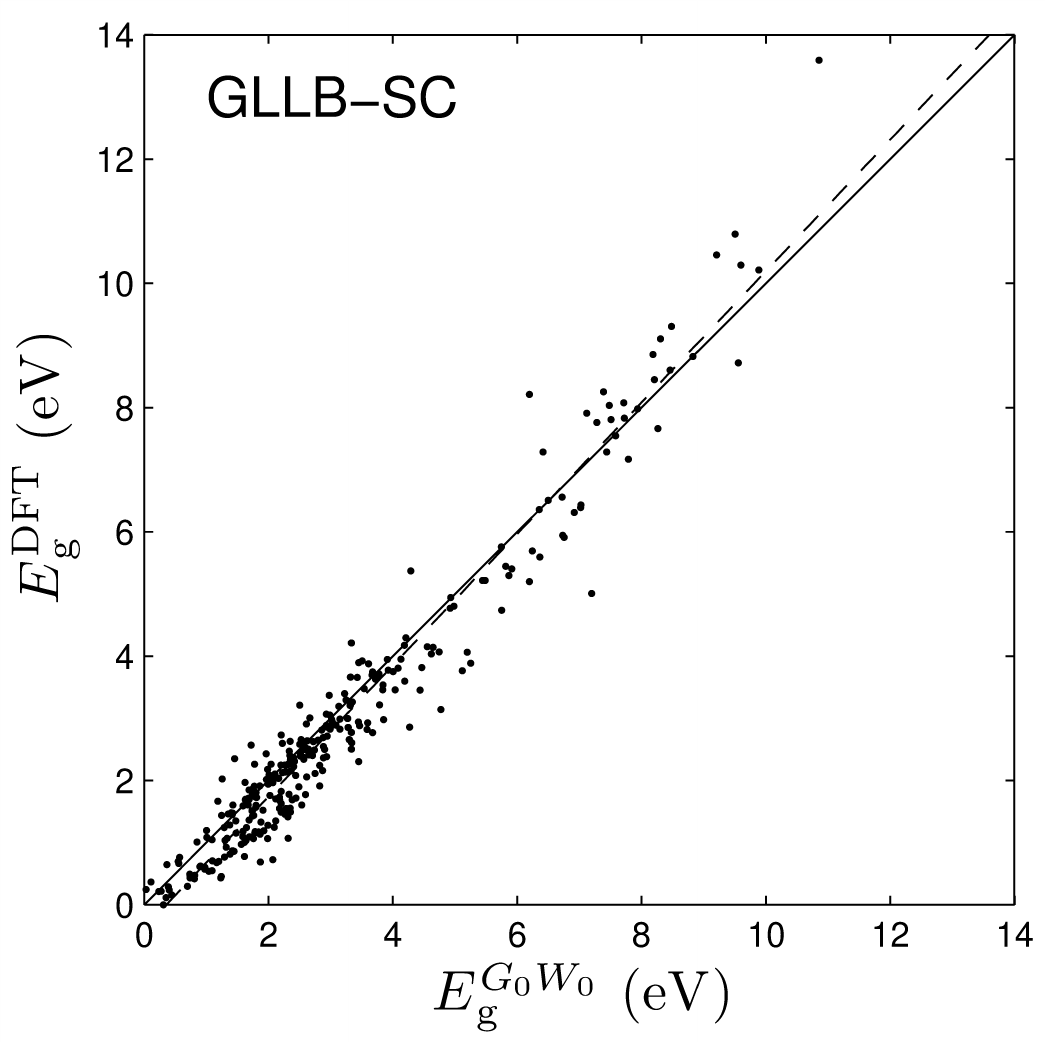}
\includegraphics[scale=0.5]{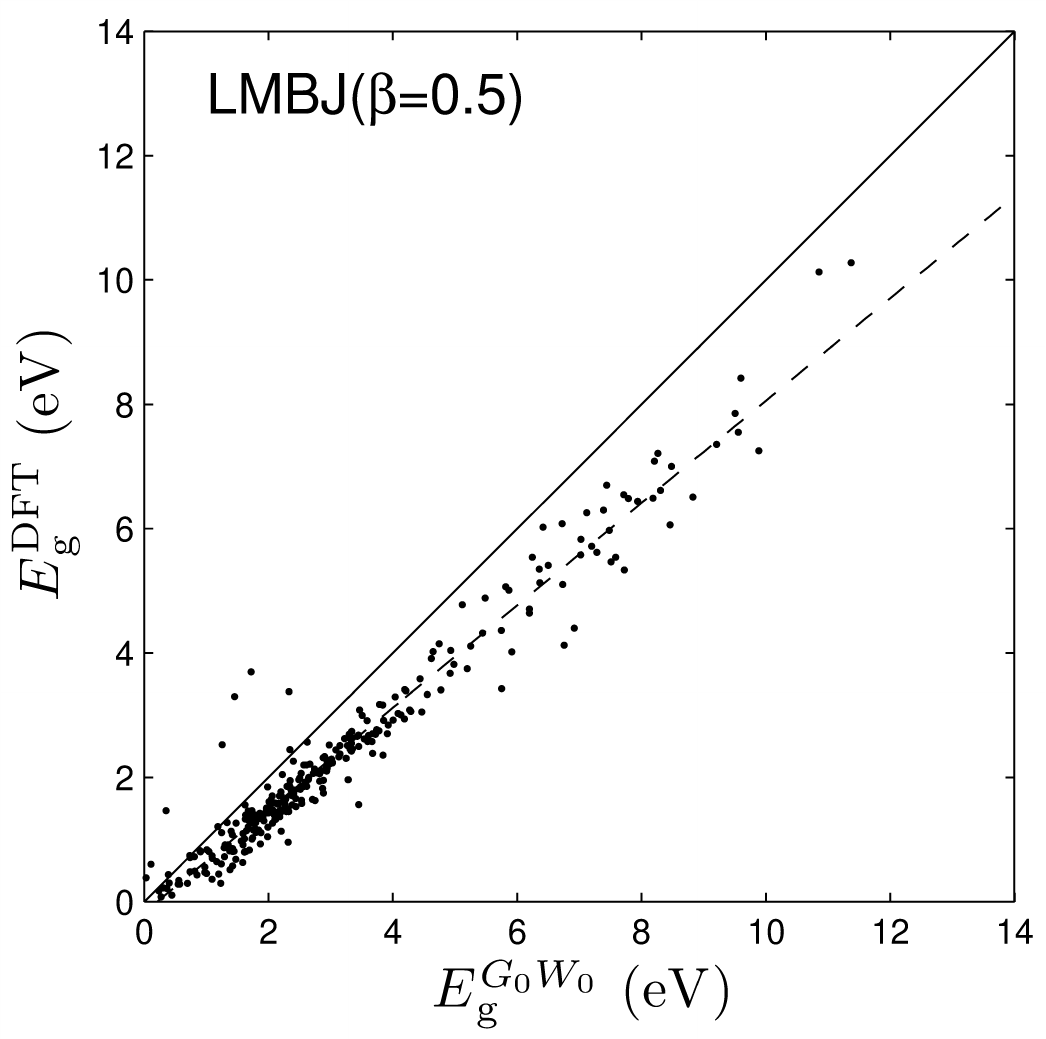}
\includegraphics[scale=0.5]{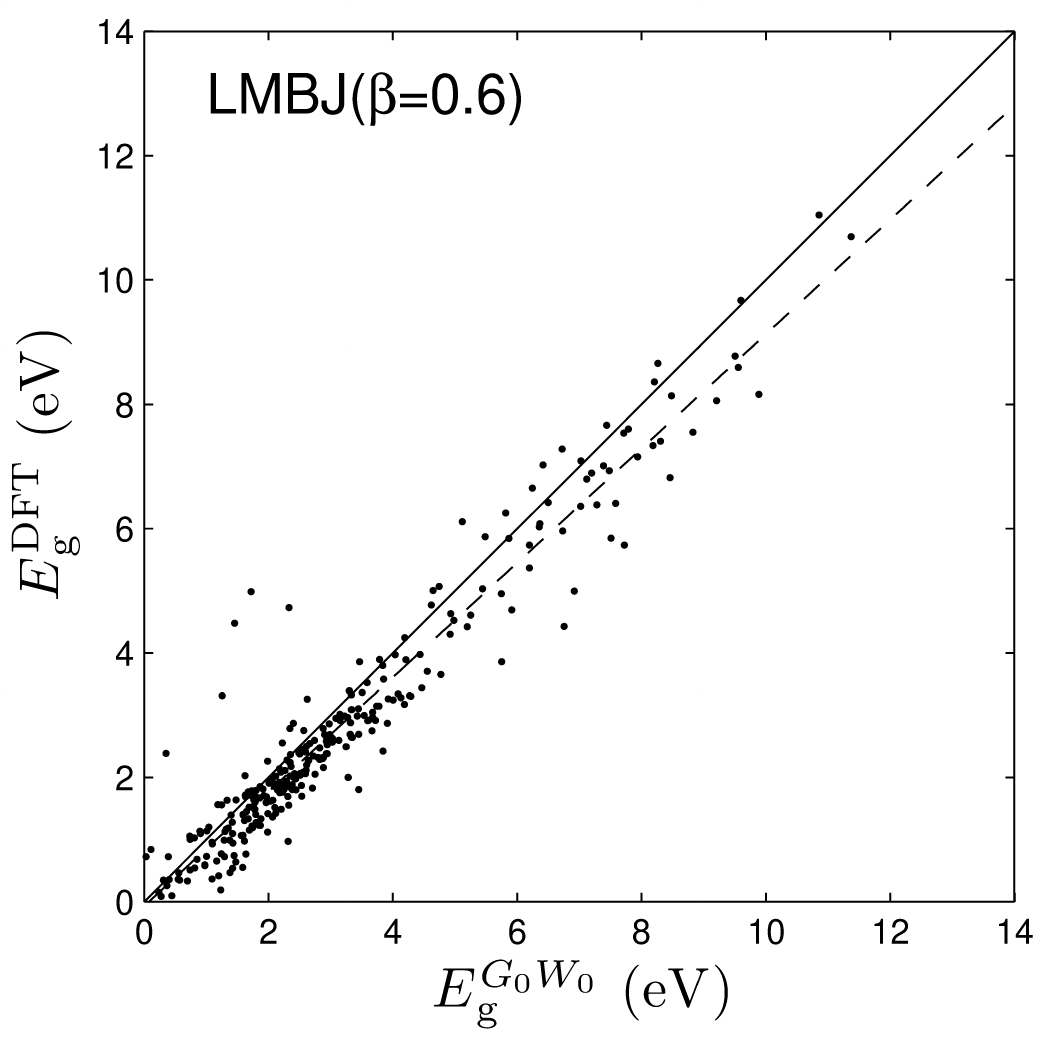}
\caption{\label{fig:band_gap}DFT versus $G_{0}W_{0}$ fundamental band gaps for the set of 298 2D materials. The results for LB94 are omitted. The HSE06 results are from Ref.~\onlinecite{Haastrup2DM18}. The dashed line represents the linear fit $y=ax+b$, where $a$ and $b$ are from Table~\ref{tab:band_gap}. The band gap of CaF$_{2}$ obtained with GLLB-SC is 14.07~eV, and therefore just not visible.}
\end{figure*}

The functionals listed in Sec.~\ref{sec:method} were used to calculate the band gap of 298 of the 2D materials in the C2DB database. As mentioned, the 298 systems were selected by Rauch \textit{et al.} \cite{RauchPRB20} and $G_{0}W_{0}$ results are used as reference. Table~\ref{tab:band_gap} shows various statistical quantities: the mean (percentage) error [M(P)E], mean absolute (percentage) error [MA(P)E], and standard (percentage) deviation [S(P)D]. The coefficients $a$ and $b$ of a linear fit as well as the Pearson correlation coefficient $r$ are also listed in Table~\ref{tab:band_gap}. For comparison purposes, the results from Ref.~\onlinecite{Haastrup2DM18} obtained with the screened hybrid functional HSE06 \cite{HeydJCP03,KrukauJCP06} using the GPAW code \cite{EnkovaaraJPCM10} are shown, as well. The numerical values of the band gaps for all materials and xc functionals can be found in the supplementary material. The values in Table~\ref{tab:band_gap} are also shown graphically on Fig.~\ref{fig:radar}.

The errors obtained with PBE are large, since the MAE is 1.50~eV and the MAPE is 51\%. The two other GGAs improve over PBE. The improvement is moderate with EV93PW91, but more visible with HLE16. The latter leads to a MAE close to 1~eV and a MAPE of 39\%. The LB94 potential leads to even worse results than PBE. Thus, the correct asymptotic behavior $-1/r$ of the LB94 potential far from nuclei does not seem to be as useful for 2D materials as it is for molecules.\cite{CasidaJCP00}

Turning to the MGGAs, r$^{2}$SCAN and HLE17 perform roughly as the GGAs EV93PW91 and HLE16 with a MAE slightly above 1~eV and a MAPE just below 40\%. The three other MGGAs are more accurate, and this is particularly the case of mTASK, which leads to a MAE of 0.52~eV and a MAPE of 21\%. Therefore, mTASK as well as TASK are more accurate than the expensive hybrid functional HSE06. However, the DFT band gaps that agree best with the $G_{0}W_{0}$ reference values are those obtained with GLLB-SC. The MAE and MAPE with GLLB-SC are as low as 0.42~eV and 21\%, respectively. In passing, we note that our MAE obtained with GLLB-SC agrees very well with the MAE of 0.38~eV reported by Haastrup \textit{et al}. \cite{Haastrup2DM18} for another subset of about 250 materials of the C2DB database.

As previously observed \cite{RauchPRB20} the performance of LMBJ with the original parameters ($\beta=0.5$) and HSE06 are pretty similar, but of course the advantage of the LMBJ potential is to be computationally more efficient and to scale better with system size. The MAE of LMBJ with $\beta=0.5$ is 0.78~eV, while the MAPE is 35\%. However, if the parameter $\beta$ in Eqs.~(\ref{eq:c}) and (\ref{eq:g}) is increased to 0.6, then the MAE and MAPE are reduced. This is especially the case for the MAE which is now 0.50~eV, similar to mTASK.

The individual results are shown graphically for most DFT functionals in the panels of Fig.~\ref{fig:band_gap}. This is a convenient way to get an idea of the S(P)D and the linear fit coefficients. From Table~\ref{tab:band_gap} we can see that the SD is the lowest (in the range 0.50$-$0.56~eV) for mTASK, GLLB-SC, TASK, and LMBJ($\beta=0.5$), and the largest (close to 1~eV) for LB94, PBE, and EV93PW91. In terms of SPD, the lowest value is 16\%, obtained with EV93PW91 (one of the worst functionals for the SD). The largest values, around 150\%, are obtained with MGGAC and LMBJ($\beta=0.6$). Concerning the linear fit, the functionals with the slope ($a$) that is the closest to 1 are GLLB-SC and LMBJ($\beta=0.6$), which is also visible from Fig.~\ref{fig:band_gap}. PBE leads to the worst value of $a$ (0.57). For the offset $b$ it is noteworthy that GLLB-SC along with LB94 lead to the values that differ the most from zero.
Regarding the correlation coefficient $r$, most functionals have a value of 0.97 or 0.98, while LMBJ($\beta=0.6$) leads to 0.96 and LB94 to 0.95. Thus, except for LB94, which is the worst functional for most quantities, the correlation coefficient does not really seem to be a useful quantity.

Concerning the outliers visible in Fig.~\ref{fig:band_gap}, we note the following. There are a couple of materials with small band gaps, more particularly FeCl$_{2}$, FeBr$_{2}$, and FeI$_{2}$, that are strongly overestimated with HSE06 and LMBJ, and to a lesser degree with MGGAC, GLLB-SC, and mTASK. Nevertheless, it is legitimate to question the accuracy of the $G_{0}W_{0}$ band gap for the three Fe$X_{2}$ systems, in particular since also HSE06 shows large deviation, which is somehow surprising. The correlation on the Fe atom may be quite strong, such that PBE is not the appropriate functional to generate the orbitals and eigenvalues for $G_{0}W_{0}$.\cite{JiangPRB10} Furthermore, it should be noted that the real ground state of those transition-metal halides, as well as of some other systems, is ferromagnetic.

The two largest band gaps, for CaF$_{2}$ and SrF$_{2}$, are largely overestimated with GLLB-SC. Concerning the number of false metals (shown in Table~\ref{tab:band_gap}), we note that there are 7 such cases with LB94, as for instance AsIn (1.69~eV with $G_{0}W_{0}$) or PtTe$_{2}$ (1.24~eV with $G_{0}W_{0}$). HLE16 leads to three false metals, HLE17 and GLLB-SC to one such case, while no false metals are obtained with the other functionals. In this respect, LB94 is the worst functional, which is expected since it leads to the strongest underestimation in the band gap. As discussed in previous works,\cite{TranJPCA17,TranPRB20} HLE16 and HLE17 have rather oscillatory potentials
and are therefore unpredictable.

By considering all results discussed so far, we can conclude that the best approximation for the calculation of the fundamental band gap of 2D materials is GLLB-SC. It is in fact the best (or nearly the best) approximation for all statistical quantities, except the SPD and coefficient $b$ of the linear fit. mTASK, which is the second best approximation, is overall rather close to GLLB-SC. LMBJ($\beta=0.6$) and TASK are also pretty accurate, and actually at least as accurate as the hybrid HSE06. However, as noted above, the SPD of LMBJ($\beta=0.6$) is very large, which is due to a large spread of the errors for the materials with small band gaps.

\begin{figure}
\includegraphics[width=\columnwidth]{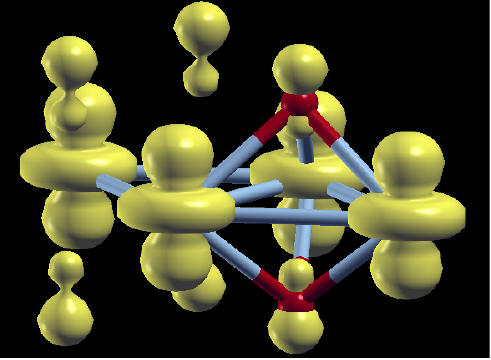}
\includegraphics[width=\columnwidth]{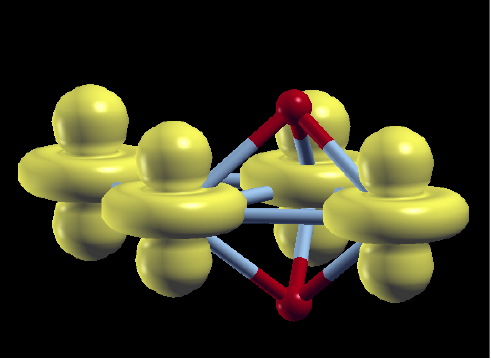}
\caption{\label{fig:CrO2}Plots of the VBM (upper panel) and CBM (lower panel) in monolayer CrO$_{2}$. The Cr-$3d_{z^{2}}$ and O-$2p_{z}$ orbitals can be recognized.}
\end{figure}

In one of our previous works on 2D materials,\cite{PatraJPCC21} plots of the xc potentials were shown to provide an explanation to some of the observed trends. For instance, for $X$S$_{2}$, $X$Se$_{2}$, and $X$Te$_{2}$, the PBE and LMBJ($\beta=0.5$) band gaps are basically the same when $X=\text{Mo}$ or W, while it is not at all the case when $X=\text{Zr}$ or Hf. We could understand these results by comparing the curves of the PBE and LMBJ potentials. Here we consider CrO$_{2}$, which, as the other 2D systems with Cr, seems to be problematic for all semilocal functionals. The $G_{0}W_{0}$ and HSE06 band gaps are 1.23 and 1.14~eV, respectively, however the semilocal methods give values that are in the range 0.19$-$0.46~eV, which is much smaller. More specifically, LMBJ($\beta=0.6$) leads to the smallest band gap (0.19~eV), while a value around 0.45~eV is obtained by PBE, EV93PW91, r$^{2}$SCAN, (m)TASK, and GLLB-SC. Figure~\ref{fig:CrO2} shows the VBM and CBM in CrO$_{2}$, where we can see that both are of Cr-$3d_{z^{2}}$ character (the VBM has also some O-$2p_{z}$ component). In such a case where the VBM and CBM are located in the same position of space and are in addition of the same character, an opening of the band gap is difficult to achieve with a multiplicative potential (a GGA or LMBJ). Actually, for CrO$_{2}$ even the non-multiplicative MGGAs and GLLB-SC can not increase the band gap with respect to PBE. Other systems with a $d$$-$$d$ gap that is underestimated with all semilocal functionals are the other isoelectronic Cr materials (e.g., CrS$_{2}$ or CrSTe), but also for instance TiCl$_{2}$ and those of the same family.

For this work, we also investigated if an improvement of the LMBJ potential as originally proposed in Ref.~\onlinecite{RauchPRB20} was possible. A full scan of the four-dimensional space of the parameters $\alpha$, $\beta$, $\sigma$, and $r_{s}^{\text{th}}$ in Eqs.~(\ref{eq:c})-(\ref{eq:g}) would be tedious. Therefore, only one or two parameters were simultaneously varied, which however should be sufficient to tell us if a clear improvement of the LMBJ accuracy is possible or not. Without going too much into detail, we observed the following. The MAE can be reduced by increasing either $\alpha$ or $\beta$, and this, as exemplified above using $\beta=0.6$, improves the results compared to the original value $\beta=0.5$. However, the MAPE was barely reduced (from 35\% to 32\%), while there was a clear increase in the SPD. The latter quantity is nearly doubled. Up to some point, increasing $\beta$ further would continue to lower the MAE and rise the MAPE, SD, and SPD. We also observed that a larger $\beta$ leads to a larger band gap for the vast majority of materials. Changing the value of $\sigma$ and/or $r_{s}^{\text{th}}$ leads to a deterioration of the results, which is rather expected since the original values were already optimized for 2D materials by Rauch \textit{et al}.\cite{RauchPRB20} $\beta=0.6$ (with $\alpha$, $\sigma$, and $r_{s}^{\text{th}}$ unmodified) is a choice among others that leads to rather well balanced errors overall if one considers the four mean errors and coefficient $a$ as the most important quantities. As discussed above, LMBJ is quite satisfying overall and does not lag far behind GLLB-SC and mTASK.

We mention that in the search of an alternative to Eq.~(\ref{eq:gbar}) for systems with vacuum, a possibility could be
\begin{equation}
\tilde{g}=\frac{1}{\tilde{V}}\int\limits_{\text{cell}}\frac{\lvert \nabla\rho(\bm{r'})\rvert}{\rho(\bm{r'})}\text{erf}\left(\frac{\rho(\bm{r}')}{\rho_{\text{th}}}\right)d^3r',
\label{eq:gtilde}
\end{equation}
where
\begin{equation}
\tilde{V}=\int\limits_{\text{cell}}\text{erf}\left(\frac{\rho(\bm{r}')}{\rho_{\text{th}}}\right)d^3r'.
\label{eq:Vtilde}
\end{equation}
In Eqs.~(\ref{eq:gtilde}) and (\ref{eq:Vtilde}), the contribution to the integral comes only from the region of space where the electron density $\rho$ has a non-negligible value. Although potentially interesting, Eq.~(\ref{eq:gtilde}) has not led to improved results. However, a more careful exploration needs to be done.

To finish this section, we mention that LMBJ has also been tested by some of us \cite{RauchJCTC21} for the ionization potential (IP) of molecules and the electronic properties of the system consisting of a F6-TCNNQ molecule adsorbed on a hydrogenated Si(111) surface. While LMBJ is accurate for the IP of molecules it is not so for the charge transfer between the molecule and the surface.

\section{\label{sec:summary}Summary}

In this work we tested a variety of xc functionals for the calculation of the band gap of 2D materials. The test set comprises 298 2D materials for which $G_{0}W_{0}$ band gaps are available and were used as reference. The tested xc functionals are the most accurate currently available for band gaps. The results show that the two most accurate are the GLLB-SC potential and the mTASK functional. The LMBJ($\beta=0.6$) potential and TASK functional can also be considered as accurate and follow quite closely GLLB-SC and mTASK.

At that point, it is also important to remind some technical aspects. GLLB-SC and LMBJ have no associated energy functional, which is inconvenient (e.g., no geometry optimization is possible and the zero-force and zero-torque conditions are not satisfied\cite{GaidukJCP09}). Furthermore, GLLB-SC depends on the eigenvalues and the Fermi energy, such that some kind of size-consistency is not satisfied. All these problems do not occur with MGGAs like TASK which consist of a energy functional. However, we should also mention that the lattice parameters of bulk solids are very inaccurate (strongly overestimated) with TASK.\cite{Doumont21} This is probably the unavoidable price to pay when a GGA or MGGA energy functional is constructed with the aim to provide very accurate band gaps.

The results presented in this work represent the most comprehensive study about the performance of semilocal xc functionals for 2D materials. They can serve as a guide for applications and future development of xc functionals. However, we believe that it will be difficult to achieve a better accuracy than GLLB-SC or mTASK, at least against the $G_{0}W_{0}$ results that we have used here as reference.

\section*{\label{supplementarymaterial}Supplementary Material}

See the supplementary material for the band gap of all materials calculated with all xc methods.

\begin{acknowledgments}

J. Doumont and P. Blaha acknowledge support from the Austrian Science Fund (FWF) through project W1243 (Solids4Fun). L. Kalantari acknowledges support from the TU-D doctoral college (TU Wien). S. Botti and P. Borlido are supported by the European Commission in the framework of the H2020 FET Open project SiLAS (GA no. 735008). M. A. L. Marques and S. Botti acknowledge partial support from the DFG though the projects TRR 227 (project B09), SFB-1375 (project A02), and BO 4280/8-1, respectively. S. Botti and T. Rauch acknowledge funding from the Volkswagen Stiftung (Momentum) through the project ``dandelion". S. Jana acknowledges funding from NISER, Bhubaneswar, India.

\end{acknowledgments}

\section*{\label{data}Data Availability}

The data that supports the findings of this study are available within the article [and its
supplementary material].

\bibliography{references}

\end{document}